\documentstyle[12pt,prd,aps,epsfig]{revtex} \tighten

\title{Lepton flavor violating $Z\rightarrow l_il_j$ in flavor-universal topcolor-assisted
technicolor }

\author{Chongxing Yue$^{(a,b)}$, Hong Li$^{b}$,  Yanming Zhang$^{b}$, Yong Jia$^{b}$ \\
{\small a: CCAST(World Laboratory) P.O. BOX 8730. B.J.
100080P.R.China}\\
{\small b: College of Physics and Information
Engineering,}\\
\small{Henan Normal University, Xinxiang  453002.
P.R.China}
\thanks{This work is supported by the National Natural Science
Foundation of China(I9905004), the Excellent Youth Foundation of
Henan Scientific Committee(9911), and Foundation of Henan
Educational Committee.}
\thanks{E-mail:cxyue@public.xxptt.ha.cn}
\thanks{Mailing address} }
\date{\today}
\begin{document}
\maketitle
\begin{abstract}
\hspace{5mm} In the context of flavor-universal topcolor-assisted
technicolor (TC2) models, we calculate the lepton flavor violating
(LFV)  $Z\rightarrow l_il_j$ decays. We find that the extra $U(1)$
gauge boson $Z^{\prime}$ can give significant contributions to
these LFV processes. With reasonable values of the parameters, the
branching ratios of the processes $Z\rightarrow \tau\mu$ and
$Z\rightarrow \tau e$ can approach the experimental upper limits.
The indirect bound on the process $Z\rightarrow \mu e$ can give a
severe constraint on the flavor-universal TC2 models.

\end {abstract}

\newpage
The high statistic results of the Superkamiokande(SK) atmospheric
neutrino experiment\cite{y1} and the solar neutrino
experiment\cite{y2} have made one to believe that neutrinos are
massive and oscillate in flavor and are of interest in the lepton
flavor violating (LFV) processes. Among them, the LFV $Z$ decays,
such as $Z\rightarrow \tau e, Z\rightarrow \tau\mu$, and
$Z\rightarrow \mu e$ are important for the search of neutrinos and
the physics beyond the standard model(SM). The Giga $Z$ option of
the TESLA linear collider project will work at the $Z$ resonance
and increase the production rate of $Z$ boson at
resonance\cite{y3}. This force one to study the LFV $Z$ decays
more precisely.

It is well known that the lepton numbers are automatically
conserved and the tree level LFV $Z$ decays are absent in the SM.
Thus, one needs an extended theory to describe the LFV $Z$ decays.
The $\nu SM$\cite{y4}, which takes neutrinos massive and permits
the lepton mixing mechanism, is one of the possibility. Ref.[5]
has studied the LFV $Z$ decays in the context of $\nu SM$.
However, the branching ratios($BR's$) are very small, i.e.,
$BR(Z\rightarrow e\mu)\sim BR(Z\rightarrow e\tau)\sim 10^{-54}$
and $BR(Z\rightarrow \mu\tau)<4\times 10^{-60}$. The $BR's$ of $Z$
decays can be increased in the framework of the $\nu SM$
 with heavy neutrino\cite{y6}. Recently, the LFV $Z$ decays are
 studied in the context of the Zee model\cite{y7} and the general
 2HDM type III \cite{y8}. Ref.[8] has shown that, with the model
 parameters in the restriction region, the $BR's$ can be highly
 enhanced which can reach $7\times 10^{-11}$ and $2.5\times
 10^{-9}$ for $BR( Z\rightarrow \mu e)$ and $BR(Z\rightarrow \tau
 e)$, respectively.

 The aim of this letter is to point out that the $BR's$ of the LFV
 $Z$ decays can be significantly enhanced in the flavor-universal
 topcolor assisted technicolor(TC2) models\cite{y9}, which may
 approach the present experimental upper limits:
\begin{eqnarray}
\nonumber BR (Z \rightarrow \tau e)<9.8\times
10^{-6}\cite{y10,y11},\\ \nonumber  BR (Z\rightarrow \tau
\mu)<1.2\times 10^{-6}\cite{y10,y12},\\  BR (Z \rightarrow \mu
e)<1.7\times 10^{-6}\cite{y10}.
\end{eqnarray}
and with the improved sensitivity at Giga-Z, these numbers could
be pulled down to \cite{y13}:
\begin{equation}
BR (Z \rightarrow \tau e)<f\times 1.5\times 10^{-8},\hspace{5mm}
BR (Z\rightarrow \tau \mu)<f\times 2.2\times
 10^{-8},\hspace{5mm} BR (Z \rightarrow \mu
 e)<2\times 10^{-9}.
\end{equation}
with $f=0.2-1.0$.

Given the large value of the top quark mass and the sizable
splitting between the masses of the top and bottom quarks, it is
natural to wonder whether $m_t$ has a different origin from the
masses of the other quarks and leptons. There may be a common
origin for electroweak symmetry breaking(EWSB) and top quark mass
generation. Much theoretical work has been carried out in
connection to the top quark and EWSB. The TC2 models\cite{y14} and
the flavor-universal TC2 models\cite{y9} are two of such examples.
All of these models suppose that top quark condensation contribute
to some of EWSB. The strong $U(1)_y$ structure is required to tilt
the chiral condensation in the $t\bar{t}$ direction and not form a
$b\bar{b}$ condensation. Thus, this kind of models generally
predict the existence of the extra $U(1)$ gauge boson
$Z^{\prime}$, which has flavor changing coupling vertices, such as
$Z^{\prime}tc$, $Z^{\prime}\tau\mu$, $Z^{\prime}\tau e$. The new
gauge boson $Z^{\prime}$ may have significant contributions to
some flavor changing neutral current processes\cite{y15}. Thus, in
this letter, we calculate the contributions of the gauge boson
$Z^{\prime}$ to the LFV $Z$ decays.

In the flavor-universal TC2 models\cite{y9}, the gauge group is
the same as in the traditional TC2 models\cite{y14}:
\begin{equation}
G_{ETC}\times SU(3)_1\times SU(3)_2\times SU(2)\times U(1)_1\times
U(1)_2.
\end{equation}
At an energy scale $\Lambda$, the color sector ($SU(3)_1\times
SU(3)_{2}$) breaks to its diagonal subgroup $SU(3)_c$ and the
hypercharge groups break in the pattern $U(1)_1\times
U(2)_2\rightarrow U(1)_y$. Thus these models also predict the
existence of two additional gauge bosons : colorons and
$Z^{\prime}$. However, the fermion charge assignments are
significantly different from those in TC2 models. In quark sector,
all quarks are $SU(3)_1$ triplets and $SU(3)_2$ singlets. In the
hypercharge sector, the third generation of fermions transforms
under the stronger $U(1)_1$ and the others transform under the
weaker $U(1)_2$. All of the quarks and leptons have the same weak
charge assignments as in the SM, which are showed in Table 1 of
Ref.[9].

The flavor-diagonal couplings of $Z^{\prime}$ to leptons can be
written as:
\begin{eqnarray}
\nonumber
\pounds_{Z^\prime}^{FD}&=&-\frac{1}{2}g_{1}\cot\theta_{y}
Z_{\mu}^\prime(\bar{\tau}_{L}\gamma^{\mu}\tau_{L}+2
\bar{\tau}_{R}\gamma^{\mu}\tau_{R})
\\
 & & +\frac{1}{2}g_{1}\tan\theta_{y}Z_{\mu}^\prime(\bar{\mu}_{L}
 \gamma^{\mu}\mu_{L}+2\bar{\mu}_{R}\gamma^{\mu}\mu_{R}+\bar{e}_{L}
 \gamma^{\mu}e_{L}+2\bar{e}_{R}\gamma^{\mu}e_{R}),
\end{eqnarray}
where $g_1$ is the ordinary hypercharge gauge coupling constant,
$\theta_y$ is the mixing angle with
$\tan\theta_y=\frac{g_1}{(2\sqrt{\pi K_1})}$. To obtain the top
quark condensation, there must be $\tan\theta_y\ll 1$. The flavor
changing couplings of $Z^{\prime}$ to leptons can be written as:
\begin{eqnarray}
 \nonumber \pounds_{Z^\prime}^{FC}&=&-\frac{1}{2}g_{1}Z_{\mu}^\prime
 [K_{\tau\mu}(\bar{\tau}_{L}\gamma^{\mu}\mu_{L}+
2\bar{\tau}_{R}\gamma^{\mu}\mu_{R})  \\
 & & +K_{\tau e}(\bar{\tau}_{L}\gamma^{\mu}e_{L}+
2\bar{\tau}_{R}\gamma^{\mu}e_{R})+K_{\mu e
}\tan^{2}\theta_{y}(\bar{\mu}_{L}\gamma^{\mu}e_{L}+
2\bar{\mu}_{R}\gamma^{\mu}e_{R})],
\end{eqnarray}
where $K_{ij}$ are the flavor mixing factors. In the following
estimation, we will assume $K_{\tau\mu}=K_{\tau e}=K_{\mu
e}=K=\lambda$\cite{y16}, where $\lambda=0.22$ is the Wolfenstein
parameter\cite{y17}.

From Eq.(4) and Eq.(5), we can see that the extra gauge boson
$Z^{\prime}$ exchange can indeed induce the LFV $Z$ decays
$Z\rightarrow l_il_j$. The relevant diagrams are depicted in
Fig.1. Similar to Ref.[18], we can calculate these diagrams. In
our calculation , we have taken $m_{\mu}\simeq 0, m_{\tau}\simeq
0$, and $m_{e}\simeq 0$. It is easy to see that, relative to the
contributions of Fig.1(a), the contributions of Fig.1(b) and
Fig.1(c) to $Z\rightarrow \tau\mu$ decay are suppressed by the
factors $\tan^2\theta_y$ and $\tan^3\theta_y$, respectively. The
conclusions are also apply to $Z\rightarrow \tau e$ and
$Z\rightarrow \mu e$ decays. Then, the LFV couplings of $Z$ arised
from the gauge boson $Z^{\prime}$ exchange can be written as:
\begin{equation}
\delta g_L^{\tau\mu}=\delta g_L^{\tau
e}\simeq\frac{K_1\tan\theta_y}{6\pi}g_L^lK[\frac{m_Z^2}{M_Z^2}\ln\frac{M_Z^2}{m_Z^2}],
\end{equation}
\begin{equation}
\delta g_R^{\tau\mu}=\delta g_R^{\tau
e}\simeq\frac{2K_1\tan\theta_y}{3\pi}g_R^lK[\frac{m_Z^2}{M_Z^2}\ln\frac{M_Z^2}{m_Z^2}],
\end{equation}
\begin{equation}
\delta g_L^{\mu
e}\simeq\frac{K_1\tan^2\theta_y}{6\pi}g_L^lK^2[\frac{m_Z^2}{M_Z^2}\ln\frac{M_Z^2}{m_Z^2}],
\end{equation}
\begin{equation}
\delta g_R^{\mu e
}\simeq\frac{2K_1\tan^2\theta_y}{3\pi}g_R^lK^2[\frac{m_Z^2}{M_Z^2}
\ln\frac{M_Z^2}{m_Z^2}],
\end{equation}
with
\begin{equation}
\nonumber g_L^l=\frac{e}{S_W C_W}(-\frac{1}{2}+S_W^2),
\hspace{10mm} g_R^l=\frac{e }{S_W C_W}(S_W^2),
\end{equation}
where $g_L^l(g_R^l)$ is the left(right)-handed $Z-l-l$ coupling
constant in the SM.

In general, the partial widths of the LFV $Z$ decays can be
written as:
\begin{equation}
\Gamma(Z\rightarrow l_il_j)=\frac{G_Fm_Z^3}{3\sqrt{2}\pi}[(\delta
g_L^{ij})^2+(\delta g_R^{ij})^2]
\end{equation}

To obtain numerical results, we take $m_Z=91.18$GeV,
$G_F=1.1664\times 10^{-5}$GeV$^{-2}$, $\Gamma_Z=2.495$GeV,
$S_W^2=0.2322$\cite{y19}. To obtain proper vacuum tilting (the
topcolor interactions only condense the top quark but not the
bottom quark), the coupling constant $K_1$ should satisfy certain
constraint, i. e. $K_1\leq 1$\cite{y9}. Ref.[20] has given the
lower bounds on the $Z^{\prime}$ mass from precision electroweak
fits, which range from $500$GeV to $2$TeV depending on the value
of $K_1$. Recently, Simmons \cite{y21} has shown that the B-meson
mixing can given lower bound. The results are $M_Z\geq 590$GeV if
ETC does not contribute to $\epsilon$ which is the CP-violation
parameter and $M_Z> 910$GeV if it does. In the following
calculation, we will take the $Z^{\prime}$ mass $M_Z$ and $K_1$ as
free parameters.

In Fig.2, we plot the branching ratios of the LFV $Z$ decay
processes $Z\rightarrow \tau\mu$, $Z\rightarrow \tau e$ as
function of the $Z^{\prime}$ mass $M_Z$ for three values of the
parameter $K_1$. From Fig.2 we can see that the branching ratios
decrease with $M_Z$ increasing and increase with $K_1$ increasing.
In most of the parameter space of the flavor-universal TC2 models,
the branching ratios of the processes $Z\rightarrow \tau\mu$ and
$Z\rightarrow \tau e$ are larger than $1\times 10^{-10}$. For
$K_1=1$ and $M_Z=500$GeV, we have $BR(Z\rightarrow \tau\mu, \tau
e)=7.0\times 10^{-9}$, which can approach the future experimental
upper limits: $BR(Z\rightarrow \tau e)<f\times 1.5\times 10^{-8}$
and $BR(Z\rightarrow \tau\mu)<f\times 2.2\times 10^{-8}$ with
$f=0.2-1.0$ \cite{y13}. We plot the branching ratio
$BR(Z\rightarrow \mu e)$ versus $M_Z$ in Fig. 3 for the three
values of the parameter $K_1$. One can see from Fig. 3 that
$BR(Z\rightarrow \mu e)$ is smaller than $BR(Z\rightarrow \tau\mu
$ or $\tau e)$. For $K_1=0.6,BR(Z\rightarrow \mu e)$ varies
between $3.8\times 10^{-10}$ and $1.2\times 10^{-11}$ for $M_Z$ in
the range of 500GeV - 1500GeV.

Using the current upper bound on the process $\mu\rightarrow 3e$
and other data pertaining to the $e^+e^-$ widths of the
electroweak gauge boson $Z$, Ref.[22] gives a set of bounds for
the LFV $Z$ decays:
\begin{equation}
BR(Z\rightarrow \mu e)\leq 5\times 10^{-13},\hspace{10mm}
BR(Z\rightarrow \tau l )\leq 3\times 10^{-6},
\end{equation}
with $l=e$ or $\mu$. From Fig.2 and Fig.3, we can see that these
bounds can not give any constraint on the flavor-universal TC2
models via the LFV processes $Z\rightarrow \tau\mu$ and
$Z\rightarrow \tau e$. However, it is not this case for the
process $Z\rightarrow \mu e$. To see how the bound
$BR(Z\rightarrow \mu e)\leq 5\times 10^{-13}$ constrains the
flavor-universal TC2 models, we give the contour line of
$BR(Z\rightarrow\mu e)=5\times 10^{-13}$ in the ($K_{\mu e}, M_Z$)
plane for $500$GeV $\leq M_Z\leq 1500$GeV and  three values of
$K_1$ in Fig.4. We can see from Fig.4 that the constraints on the
flavor mixing factor $K_{\mu e}$ from the indirect bound
$BR(Z\rightarrow \mu e)\leq 5\times 10^{-13}$ are very strong. If
we take $K_1=0.2$, $ M_Z\leq 1500$GeV, there must be $K_{\mu
e}\leq 0.17$.

Similar to the flavor-universal TC2 models, TC2 models predict the
existence of an extra $U(1)_y$ gauge boson $Z^{\prime}$, which can
also induce the tree-level flavor changing couplings. Thus, the
$Z^{\prime}$ can also give contributions to the LFV $Z$ decays
$Z\rightarrow l_il_j$. However, Ref.[23] has shown that $B\bar{B}$
mixing provides lower bounds on the mass of the gauge boson
$Z^{\prime}$ predicted by TC2 models, i. e., it must be larger
than 4TeV. So the contributions of the TC2 models to the LFV $Z$
decays are much smaller than those of the flavor-universal TC2
models. In most of the parameter space of the TC2 models, we have
$BR(Z\rightarrow\tau\mu)=BR(Z\rightarrow\tau e)<1\times 10^{-11}$
and $BR(Z\rightarrow\mu e)<1\times 10^{-13}$.

To completely avoid the problems, such as triviality and
unnaturalness arising from the elementary Higgs in the SM, various
kinds of dynamical EWSB theories have been proposed and among
which the strong top dynamical EWSB theories are very interesting.
This type of models generally predict the existence of an extra
$U(1)_y$ gauge boson $Z^{\prime}$ which can induce the flavor
changing coupling vertices, such as $Z^{\prime}\tau\mu,
Z^{\prime}\tau e, Z^{\prime}\mu e$. Thus, the gauge boson
$Z^{\prime}$ may have significant contributions to the LFV $Z$
decays $Z\rightarrow l_il_j$. In this letter, we calculate the
contributions of the $Z^{\prime}$ predicted by the
flavor-universal TC2 models to these decay processes. We find that
the branching ratios of the LFV $Z$ decays indeed can be
significantly enhanced by the $Z^{\prime}$ exchange. With
reasonable values of the parameters, the branching ratios of the
processes $Z\rightarrow\tau\mu$ and $Z\rightarrow\tau e$ can reach
$1\times 10^{-8}$ which can approach the future experimental upper
limits. We further consider whether the indirect bound of the
process $Z\rightarrow\mu e$ given by Ref. [21] can give severe
constraint on the flavor-universal TC2 models. Our results show
that it is indeed this case. For $K_1=1$, there must be $M_Z\geq
5.2 $TeV for $K_{\mu e}=0.22$ and $K_{\mu e}\leq 0.038$ for
$M_Z=500$GeV.

\newpage
\begin{center}
{\bf Figure captions}
\end{center}
\begin{description}
\item[Fig.1:]The diagrams of $Z\rightarrow\tau\mu$ decay due to $Z^{\prime}$
 exchange in the flavor-universal TC2 models. The diagrams of
 $Z\rightarrow\tau e$ and $Z\rightarrow\mu e$ decays can be
 obtained by appropriate replacement of the internal fermion.
\item[Fig.2:]The branching ratios of the LFV $Z$ decays $Z\rightarrow
 \tau\mu$
and $Z\rightarrow \tau e$ as function of $M_Z$.
\item[Fig.3:]The branching ratios of the LFV $Z$ decay $Z\rightarrow \mu e$
 as a function of $M_Z$.
\item[Fig.4:]The contour line of $BR ( Z\rightarrow \mu e )=5\times
10^{-13}$ in the ($K_{\mu e}, M_Z$) plane for $K_1=0.1$ \\  (
solid line ), 0.5 ( dashed line ) and 1 ( dotted line ).
\end{description}

\newpage
\begin{figure}[pt]
\begin{center}
\begin{picture}(250,200)(0,0)
\put(20,-180){\epsfxsize120mm\epsfbox{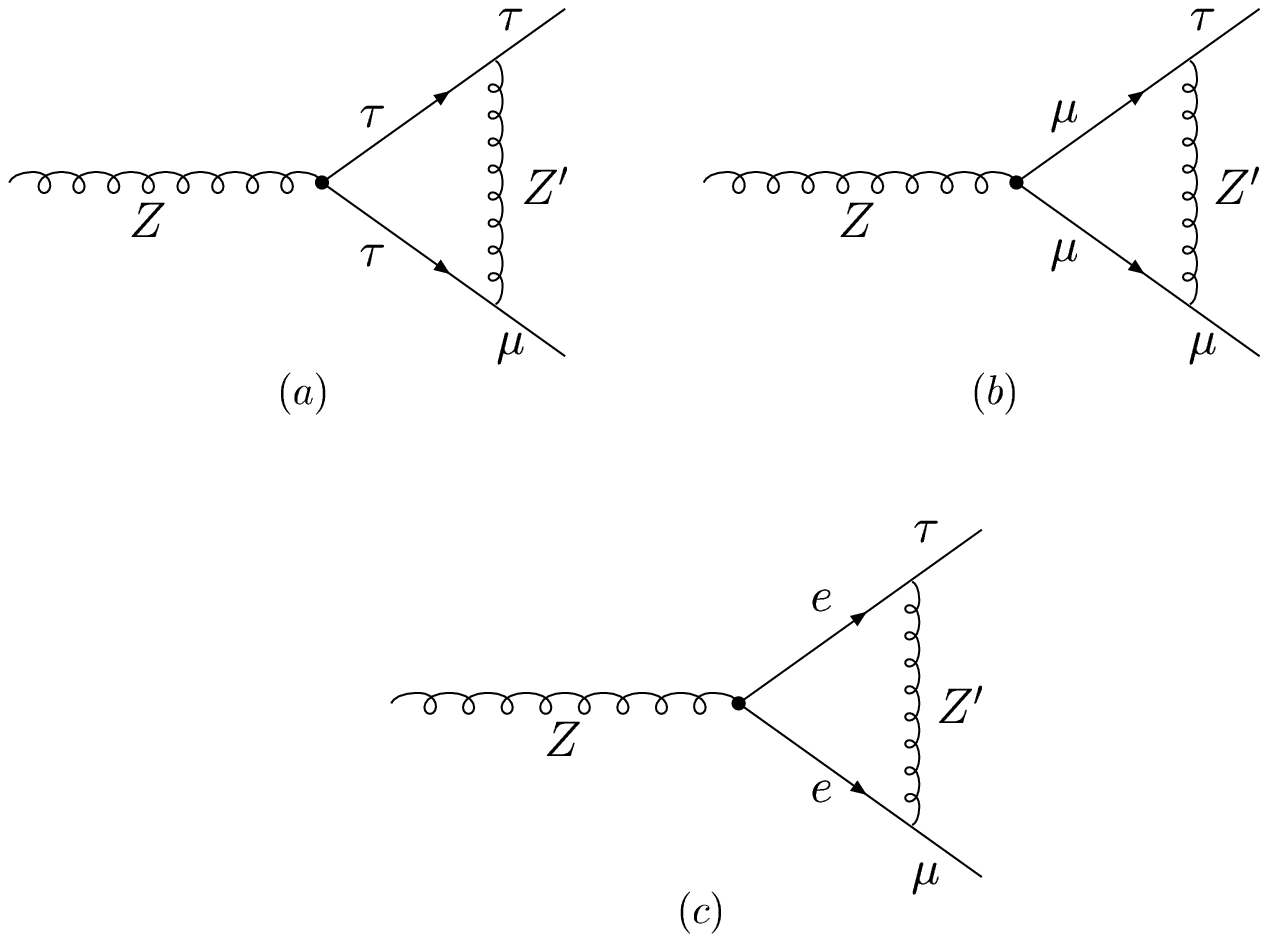}}
 \put(180,-10){Fig.1}
\end{picture}
\end{center}
\end{figure}

\newpage
\begin{figure}[b]
\begin{center}
\epsfig{file=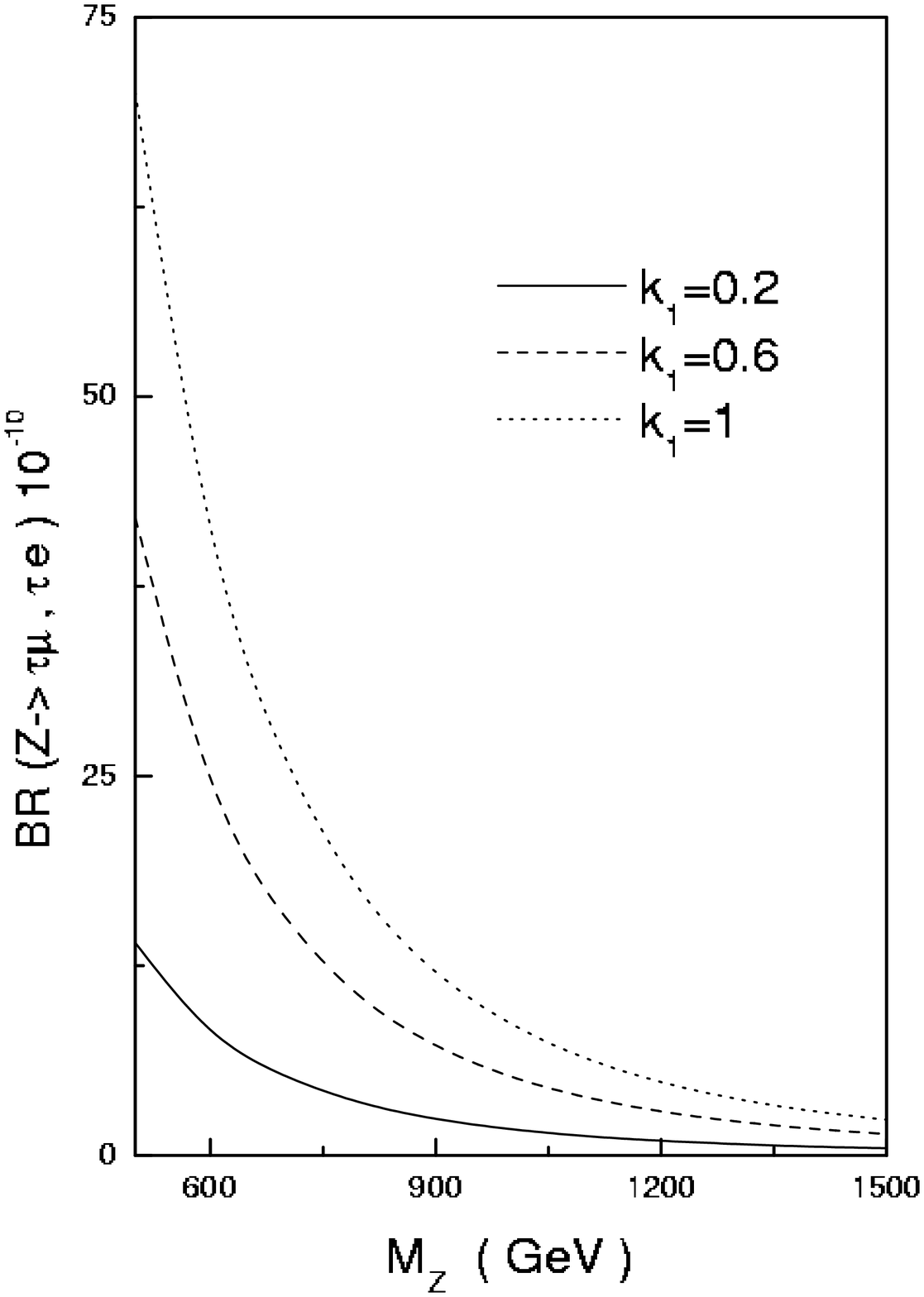,width=370pt,height=275pt}
\end{center}
\end{figure}
\begin{center} {Fig.2}
\end{center}

\newpage
\begin{figure}[hb]
\begin{center} \epsfig{file=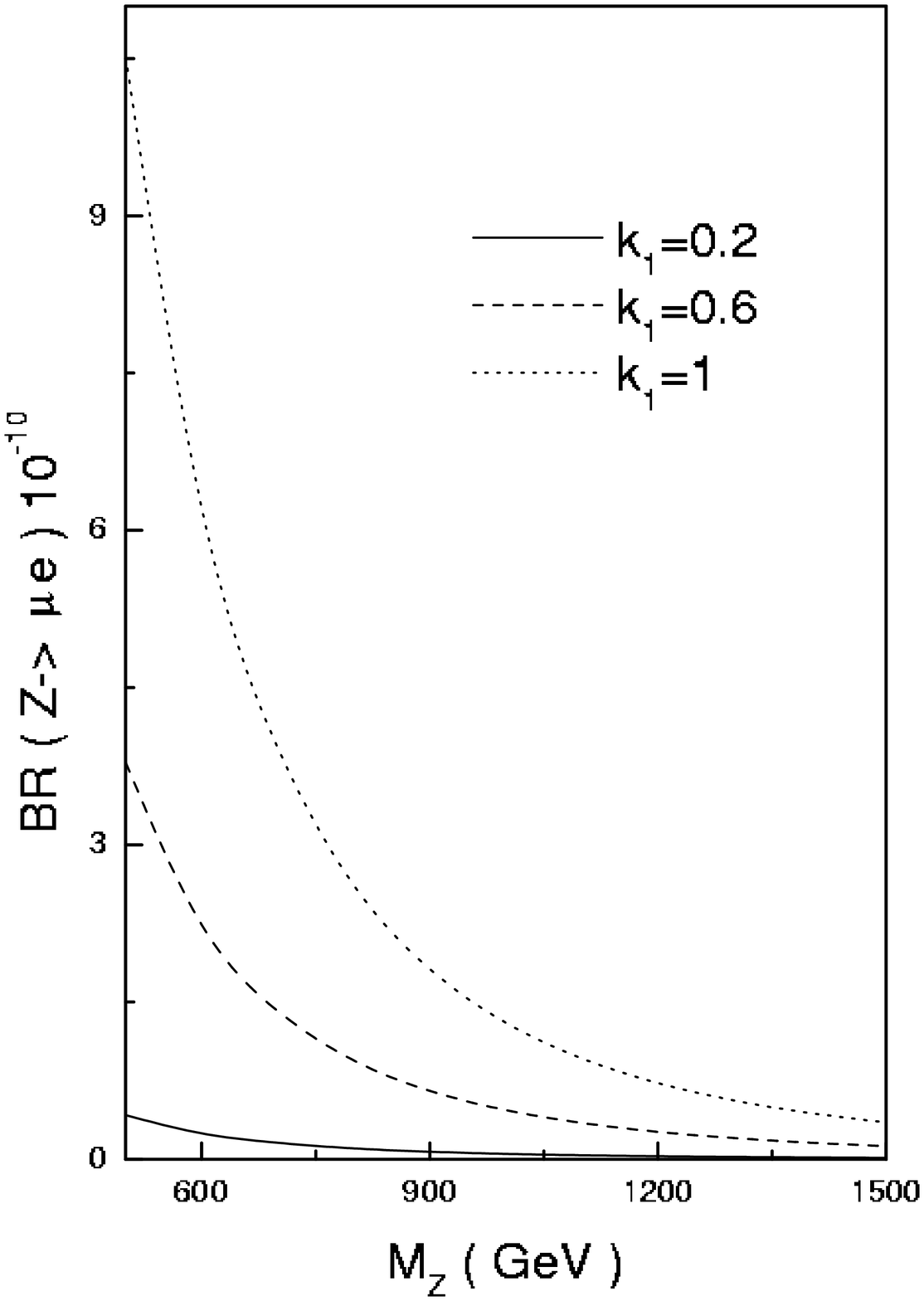,width=370pt,height=275pt}
\end{center}
\end{figure}
\begin{center}Fig.3
\end{center}

\newpage
\begin{figure}[hb]
\begin{center} \epsfig{file=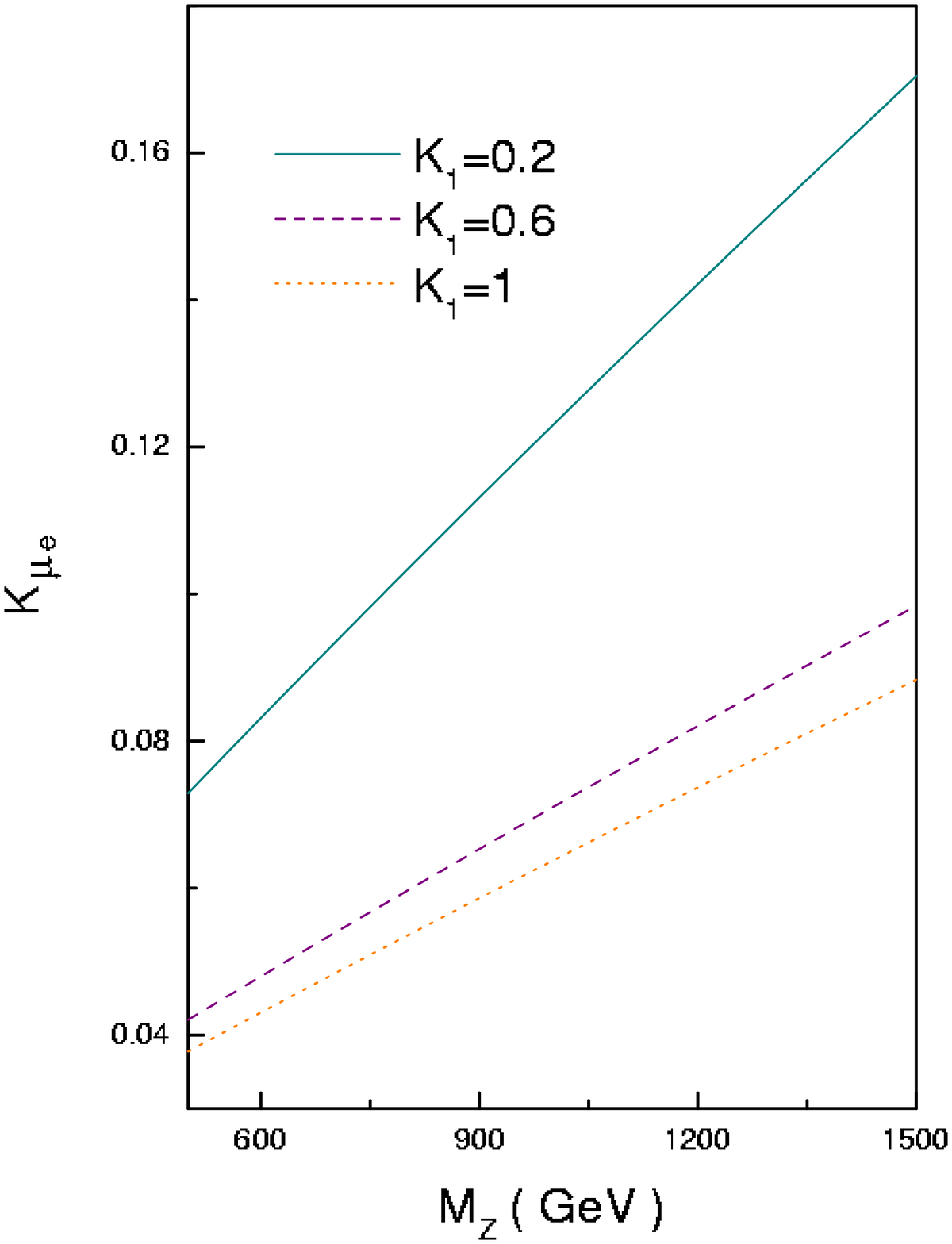,width=370pt,height=275pt}
\end{center}
\end{figure}
\begin{center}Fig.4
\end{center}


\newpage
\begin{thebibliography}{99}
\bibitem{y1}Superkamiokande
Collaboration, Y. Fukuda et al. , {\em Phys. Lett. B}{\bf335}
(1994)237; Y. Fukuda et al. , {\em Phys. Rev. Lett}. {\bf 81}
(1998)1562; {\em Phys. Rev. Lett}. {\bf 86} (2001)5651; {\em Phys.
Rev. Lett}. {\bf 86} (2001)5656.
\bibitem{y2}SNO Collaboration, Q. R. Ahmad et al. , {\em Phys. Rev. Lett}.
 {\bf 87}(2001)071301.
\bibitem{y3}R. Hawkings and K. Monig, {\em Eur. Phys. J. direct} {\bf  C8}
(1999)1.
\bibitem{y4}B. Pontecorvo, {\em Zh. Eksp. Theor. Fiz. } {\bf 33}
(1957)549; Z. Maki, M. Nakagawa and S. Sakata, {\em Prog. Theor.
Phys.} {\bf 28} (1962)870; B. Pontecorvo, {\em Sov. Phys. } JETP
{\bf 26} (1968)984.
\bibitem{y5}T. Riemann and G. Mann,  ``Nondiagonal Z decay: $Z\rightarrow
e\mu$", in Proc. of the Int. Conf. Neutrino'82, 14-19 June 1982,
Balatonfured, Hungary (A. Frenkel and E. Jenik, eds), Vol. II, PP.
58, Budapest, 1982; V. Ganapathi, T. Weiler, E. Laermann, I.
Schmitt, and P. Zerwas, {\em Phys. Rev. D}{\bf 27} (1983)579; M.
Clements, C. Footman, A. Kronfeld, S. Narasimhan, and D.
Photiadis, {\em Phys. Rev. D}{\bf 27} (1983)570; G. Mann and T.
Riemann, {\em Annalen Phys.} {\bf 40} (1984)334.
\bibitem{y6}J. I. Illana, M. Jack and T. Riemann, hep-ph/0001273;
J. I. Illana and
 T. Riemann, {\em Phys. Rev. D}{\bf 63} (2001)053004.
\bibitem{y7}A. Ghosal, Y. Koide, and H. Fusaoka, {\em Phys. Rev. D} {\bf
64} (2001)053012.
\bibitem{y8}E. O. Iltan and I. Turan, {\em Phys. Rev. D} {\bf
65} (2002)013001.
\bibitem{y9}M. B. Popovic, E. H. Simmons, {\em Phys. Rev. D} {\bf
58} (1998)095007; G. Burdman, N. Evans, {\em Phys. Rev. D} {\bf
59} (1999)115005.
\bibitem{y10}OPAL Collaboration, R. Akers et al., {\em Z. Phys. }{\bf C67}
(1995)555.
\bibitem{y11}L3 Collaboration, O. Adriani et al., {\em Phys. Lett.
              B}{\bf 316} (1993)427.
\bibitem{y12}DELPHI Collaboration, P. Abreu et al., {\em Z. Phys. }{\bf C73}
(1997)243.
\bibitem{y13}G. Wilson, ``Neutrino Oscillations: are lepton-flavor
violating Z decays ovservable with the CDR detector?" and  "Update
on experimental aspects of lepton-flavor violation", talks at
DESY-ECFA LC Workshop held at Frascati, Nov. 1998 and at Oxford,
march 1999.
\bibitem{y14}C. T. Hill, {\em Phys. Lett. B}{\bf 345} (1995)483;
K. Lane and E. Eichten, {\em Phys. Lett. B}{\bf 352} (1995)383;
 K. Lane, {\em Phys. Lett. B}{\bf 433} (1998)96 G. Cvetic,
 {\em Rev. Mod. Phys.} {\bf 71} (1999)513.
\bibitem{y15}T. Rador, {\em Phys. Rev. D}{\bf 60} (1999)095012; Chongxing Yue, et al.,
{\em Phys. Lett. B}{\bf 496} (2000)89.
\bibitem{y16}G. Buchalla, G. Burdman, C. T. Hill, D. Kominis,
{\em Phys. Rev. D} {\bf 53} (1996)5185.
\bibitem{y17}L. Wolfenstein, {\em Phys. Rev. Lett.} {\bf 51}
(1983)1945.
\bibitem{y18}C. T. Hill and Xinmin Zhang, {\em Phys. Rev. D} {\bf 51}
(1995)3563.
\bibitem{y19}Paratical Data Group, {\em Eur. Phys. J.} {\bf
             C15} (2000)1.
\bibitem{y20}R. S. Chivukula, E. H. Simmons and J. Terning, {\em Phys. Lett. B.}
{\bf 385} (1996)209.
\bibitem{y21}E. H. Simmons, {\em Phys. Lett. B.}
{\bf 526} (2002)365.
\bibitem{y22}S. Nussinov, R. D. Peccei and X. M. Zhang, {\em Phys. Rev. D.} {\bf 63}
(2000)016003; D. Delepine and F. Vissani, {\em Phys. Lett. B.}
{\bf 522} (2001)95.
\bibitem{y23}G. Burdman, K. Lane, T. Rador, {\em Phys. Lett. B}
{\bf 514} (2001)41.
\end{thebibliography}
\end{document}